\documentstyle[11pt,epsfig]{article}
\title{\bf Brane-world black hole entropy from  modified dispersion relations}
\author{A. S. Sefiedgar\thanks{e-mail: a-sefiedgar@sbu.ac.ir} and H. R. Sepangi
\thanks{email: hr-sepangi@sbu.ac.ir}
\\ {\small Department of Physics, Shahid Beheshti University, G. C.,  Evin,
Tehran 19839, Iran}}
\begin{document}
\maketitle
\begin{abstract}

The entropy of the Reissner-N\"{o}rdstrom black hole is studied
within the context of a brane-world scenario. Such a black hole is a
solution of the Einstein field equations on the brane, possessing a
tidal charge which is a reflection of the extra dimension. We use
the modified dispersion relation to obtain the entropy of such
brane-world black holes. The resulting entropy differs from that of
the standard Bekenstein-Hawking's and contains information on the
extra dimension.
\end{abstract}
\vspace{2cm}
\section{Introduction}
A common feature of all promising candidates for quantum gravity is
the existence of a minimal observable length \cite{1,2,3,4,5}. The
modified dispersion relation (MDR) is one of the approaches
incorporating such a finite resolution of the space-time in the
theoretical framework of the standard model. MDR is a common feature
in all candidates of quantum gravity. In particular, in the study of
loop quantum gravity (LQG) and of models based on non-commutative
geometry, there has been strong interest in modifications to the
energy-momentum dispersion relation \cite{6,7,8,9,10}. Since black
holes are suitable examples of an extreme quantum gravity regime,
using MDR to study their thermodynamical behavior and comparing the
results with other approaches may further our understanding of their
properties and structure.

There are, presumably, other more reliable theories such as string
theory and loop quantum gravity with which to study black hole
thermodynamics. One may therefore use the results of such studies to
impose constraints on the MDR \cite{9,10,11} which would ultimately
result in a better insight into quantum gravity. A study along these
lines was performed in a previous work \cite{11}. In this paper, we
use the form of MDR obtained in the above mentioned study where
terms proportional to odd powers of energy are not present in our
modified dispersion relation \cite{11} and concentrate on the
brane-world black hole entropy. The motivation behind the study of
brane-world black holes stem from the fact that since the advent of
theories with extra non-compact dimensions, great progress has been
made in describing some hitherto unexplained problems in particle
physics, e. g. the hierarchy problem, without appealing to
supersymmetry \cite{12}. In such models, our physical universe is a
3-dimensional brane embedded in a higher dimensional bulk (usually
one extra dimension) where the standard matter is confined to the
brane except gravity which can propagate into the bulk
\cite{12,13,15} as well as on the brane and the size of the extra
dimensions can be much larger than the Planck length scale
\cite{12}. The large size of the extra dimension is the key for
providing a unification scale of the order of a few TeV \cite{16}.
It is therefore plausible to think of the possibility of TeV-sized
black holes being produced in the universe. Density perturbations
and phase transition in the early universe may lead to such black
holes. High energy collision processes in cosmic rays and at future
colliders such as LHC can also produce these TeV-sized black holes
\cite{17,18,19}. In such processes, matter on the 3-brane may
collapse under gravity to form a black hole. To preserve the general
relativity observational predictions, the metric on the brane should
be close to the Schwarzschild metric at astrophysical scales
\cite{20}.

Brane-world black hole solutions have been studied by a number of
authors \cite{20, 21, 211}.  Of particular interest is the solution
presented in \cite{20} which is akin to that of the
Reissner-N\"{o}rdstrom solution, but without the electric charge
being present. Instead, the Reissner-N\"{o}rdstrom  type correction
to the Schwarzschild potential can be thought of as a tidal charge,
arising from the projection onto the brane of the free gravitational
field effects in the bulk. The study of entropy of  brane-world
black holes may bring about information on the extra dimensions
which, in turn, would provide a deeper insight into the quantum
theory of gravity. The entropy of a brane-world black hole was first
studied in \cite{22} within the generalized uncertainty principle
(GUP) formalism and later in \cite{23}. The entropy of a TeV-sized
Reissner-N\"{o}rdstrom type black hole within the context of the ADD
brane-world scenario was considered in \cite{16}.

In this paper, we will apply the MDR formalism to obtain the entropy
of a Reissner-N\"{o}rdstrom type black hole obtained in \cite{20}.
Such a study would be of interest and complementary to what has been
done in \cite{16} in that the results of the two approaches, namely
GUP and MDR can be compared and interpreted. This is what we intend
to do in what follows.

\section{The modified dispersion relation} The modified dispersion
relation  can be written as \cite{9}
\begin{equation}\label{1}
(\overrightarrow{p})^2=f(E,m;L_p)\simeq E^2-\mu^2+\alpha_1 L_p
E^3+\alpha_2L_p^2E^4+{\cal O}\left(L_P^3E^5\right),
\end{equation}
where $f$ is the function that gives the exact dispersion relation
and $L_p$ is the Planck length. On the right hand side we have
assumed the applicability of a Taylor-series expansion for $E\ll
\frac{1}{L_p}$. The coefficients $\alpha_i$ may take different
values in different quantum gravity approaches. Note that $m$ is the
rest energy of the particle and the mass parameter $\mu$ on the
right hand side is directly related to the rest energy, but $\mu\neq
m$ if $\alpha_i$'s do not all vanish.

Although MDR is a feature of all quantum gravity scenarios, its
functional form depends on the quantum gravity model being used. To
incorporate quantum gravitational effects, the Bekenstein-Hawking
formalism of black hole thermodynamics needs to be modified. Of
course, MDR may provide a perturbation framework for such a
modification. On the other hand, loop quantum gravity and string
theory give the entropy-area relation of black holes (for $A\gg
L_P^2$)
\begin{equation}\label{2}
S=\frac {A}{4L_P^2}+\rho \ln{\frac{A}{L_p^2}}+{\cal
O}\left(\frac{L_p^2}{A}\right),
\end{equation}
where $\rho$ may have different values in string theory and in loop
quantum gravity \cite{9,10,24}. Since string theory and loop quantum
gravity are expected to provide a more reliable solution to black
hole thermodynamics, these solutions could be considered as a test
bed against which other solutions including the ones obtained using
MDR \cite{11} should be compared. With that in mind, the entropy of
a black hole obtained using equation (\ref{1}) is functionally
different from what one obtains using string theory and loop quantum
gravity given by equation (\ref{2}). It is then necessary to
introduce constraints on the usual form of the MDR to obtain a
consistent black hole thermodynamics in both approaches. The result
is that terms proportional to odd powers of energy should be ignored
in the MDR formula \cite{11}. Consequently, we take the MDR as
\begin{equation}\label{3}
(\overrightarrow{p})^2=f(E,m;L_p)\simeq E^2-\mu^2+\alpha
L_p^2E^4+{\cal O}\left(L_P^4E^6\right),
\end{equation}
in what follows. Of course, the black hole thermodynamics obtained
via equation (\ref{3}) is now consistent with the result given by
equation (\ref{2}).

\section{Black holes on the brane} The authors in \cite{20},
working in the framework of  the Randall-Sundrum scenario, present
an exact localized black hole solution which resembles that of a
Reissner-N\"{o}rdstrom solution, but without the electric charge.
Instead, the Reissner-N\"{o}rdstrom type correction to the
Schwarzschild potential can be thought of as a tidal charge, arising
from the projection  of the free gravitational field effects in the
bulk onto the brane. These effects are transmitted via the bulk Weyl
tensor. The Schwarzschild potential $\phi=-M/(M_p^2 r)$ where $M_p$
is the effective Planck mass on the brane is modified to
\begin{equation}\label{4}
\phi=-\frac{M}{M_p^2 r}+\frac{Q}{2r^2},
\end{equation}
where $Q$ is a tidal charge parameter which may be positive or
negative. They showed that an exact black hole solution of the
effective field equations on the brane is given by the induced
metric
\begin{equation}\label{5}
ds_4^2=-fdt^2+f^{-1}dr^2+r^2d\Omega^2,
\end{equation}
where
$$f=1-\left(\frac{2M}{M_p^2}\right)\frac{1}{r}+\left(\frac{q}{\widetilde{M}_p^2}\right)\frac{1}{r^2}.$$
Note that $\widetilde{M}_p$ and $M_p$ are fundamental Planck scale
in the bulk and effective scale on the brane respectively. Since
$q=Q\widetilde{M}_p^2$ is a dimensionless tidal charge parameter
arising from the projection of gravitational field in the bulk  onto
the brane, it carries information relating to the extra dimension.
For $q>0$, this metric is analogous to the Reissner-N\"{o}rdstrom
solution with two horizons. Both of these horizons lie inside the
Schwarzschild horizon. For $q<0$, the metric has only one horizon
which is larger than the Schwarzschild horizon and is given by
$$r_+=\frac{M}{M_p^2}\left(1+\sqrt{1-\frac{qM_P^4}{M^2\widetilde{M}_P^2}}\right).$$
The negative tidal charge increases the entropy and is therefore
considered physically more natural \cite{20}.  We will concentrate
our attention on $q<0$ and denote the negative tidal charge by $q'$.
Now the metric on the $4D$ brane can be written as
\begin{equation}\label{6}
f=1-\frac{2M}{r}-\frac{q'}{r^2},
\end{equation}
with the roots given by
$$r_\pm=M\left(1\pm\sqrt{1+\frac{q'}{M^2}}\right).$$ We note that $r_+$ is the
black hole horizon and $r_-$ is negative and without physical
meaning.

\section{Entropy of a brane-world black hole in MDR formalism }
In this section we derive the entropy of a black hole within the MDR
and standard uncertainty principle. Differentiating equation
(\ref{3}) and neglecting the rest mass, we  write
\begin{equation}\label{7}
dp=dE\left(1+\frac{3}{2}\alpha
L_p^2E^2-\frac{5}{8}\alpha^2L_P^4E^4\right).
\end{equation}
One may then wrire
\begin{equation}\label{8}
dE=dp\left(1-\frac{3}{2} \alpha L_p^2E^2+\frac{23}{8}\alpha^2
L_p^4E^4\right),
\end{equation}
where we have only considered terms up to the fourth power of the
Plank length. Using the standard uncertainty principle, we  have
\begin{equation}\label{8.5}
dE\delta x\geq 1-\frac{3}{2} \alpha \frac{L_p^2}{\delta
x^2}+\frac{23}{8}\alpha^2 \frac{L_p^4}{\delta x^4}.
\end{equation}

It is interesting to note that in quantum field theory, the relation
between particle localization and its energy is given by
$E\geq\frac{1}{\delta x}$ where $\delta x$ is the particle position
uncertainty. Certainly, this relation is modified using MDR formula.
Assuming $\delta E\sim E$, we have
\begin{equation}\label{9}
E\delta x\geq 1-\frac{3}{2} \alpha \frac{L_p^2}{\delta
x^2}+\frac{23}{8}\alpha^2 \frac{L_p^4}{\delta x^4}.
\end{equation}
When a black hole is absorbing a classical particle of energy $E$
and size $R$, the minimum increase in the horizon area can be
expressed according to the following general relativistic result
\cite{25}
\begin{equation}\label{10}
(\Delta A_d)_{min}\geq\frac{8\pi L_p^{d-2} E R}{(d-3)}.
\end{equation}
In 4-dimensions, we have $(\Delta A_4)_{min}\geq{8\pi L_p^{2} E R}$.
Note that $R$ can never be smaller than $\delta x$. For convenience we may write the horizon shift as
\begin{equation}\label{11}
(\Delta A_4)_{min}\simeq{\epsilon L_p^{2} E \delta x},
\end{equation}
where $\epsilon$ is a parameter to be determined. Particles with a
Compton wavelength on the order of the inverse surface gravity are
of interest to us \cite{26,27}. Hence we can take $\delta x$ as
\begin{equation}\label{12}
\delta x\sim\kappa^{-1}=\frac{2r_+^2}{r_+-r_-}.
\end{equation}
For small  $q'$ we have
\begin{equation}\label{13}
\delta x\sim\sqrt{\frac{A}{\pi}}\left(1-\frac{4\pi q'}{A}\right),
\end{equation}
where $A=16\pi M^2$ is the outer horizon area of the black hole.
According to information theory \cite{28}, the minimum increase of a
black hole entropy is simply one bit of information which may be
presented by $b$. To obtain the entropy of the brane-world black
hole, we write
\begin{equation}\label{14}
\frac{dS}{dA}\simeq\frac{(\Delta S_4)_{min}}{(\Delta
A_4)_{min}}\simeq\frac{b}{\epsilon L_p^2 E \delta x}.
\end{equation}
Note that $L_p$ is the Planck length on the brane. Assuming that the
dimensionless ratio $L_p^2/(\delta x)^2$ is small relative to unity,
we can apply a Taylor expansion of the quantity $E \delta x$ in
equation (\ref{9}) to find
\begin{equation}\label{15}
\frac{dS}{dA}\simeq \frac{b}{\epsilon
L_p^2}\left(1+\frac{3}{2}\alpha \frac{L_p^2}{(\delta x)^2}-
\frac{5}{8}\alpha^2 \frac{L_p^4}{(\delta x)^4}+\cdots\right).
\end{equation}
For $q'\ll A$ we have
\begin{equation}\label{15.5}
\frac{dS}{dA}\simeq \frac{1}{4 L_p^2}+\frac{3}{8}\alpha \pi \left(
\frac{1}{A}+ \frac{8 \pi q'}{A^2}+\frac{48 {\pi}^2
q'^2}{A^3}\right)-\frac{5}{32}\alpha^2 L_p^2 {\pi}^2
\left(\frac{1}{A}+ \frac{8\pi q'}{A^2}+\frac{48\pi^2
q'^2}{A^3}\right)^2+\cdots.
\end{equation}
We note that setting $b/\epsilon=1/4$, the Bekenstein-Hawking area
law can be reproduced. Integrating equation (\ref{15.5}), we find
\begin{eqnarray}\label{16}
S\simeq \frac{A}{4 L_p^2}+\frac{3}{8}\alpha \pi
\ln\left(\frac{A}{4L_p^2}\right)- \frac{3}{8}\alpha \pi\left(\frac{8
\pi q'}{A}+\frac{24 {\pi}^2 q'^2}{A^2}\right)\nonumber \\+
\frac{5}{32}\alpha^2 L_p^2 {\pi}^2
\frac{1}{A}+\frac{5}{32}\alpha^2L_p^2 \pi^2\left(\frac{8\pi
q'}{A^2}+ \frac{160\pi^2 q'^2}{3A^3}\right)+\cdots+C,
\end{eqnarray}
where $C$ is an integration constant and terms smaller than
$\frac{q'^2}{A^2}$ have been neglected. To continue, we ignore $C$
and higher order correction terms in equation (\ref{16}), obtaining
\begin{equation}\label{17}
S\simeq \frac{A}{4 L_p^2}+\frac{3}{8}\alpha \pi
\ln\left(\frac{A}{4L_p^2}\right)- \frac{3}{8}\alpha \pi\left(\frac{8
\pi q'}{A}+\frac{24 {\pi}^2 q'^2}{A^2}\right)+ \frac{5}{32}\alpha^2
L_p^2 {\pi}^2 \frac{1}{A}+\frac{5}{32}\alpha^2L_p^2
\pi^2\left(\frac{8\pi q'}{A^2}+ \frac{160\pi^2 q'^2}{3A^3}\right).
\end{equation}
Of course, it goes without saying that taking into account higher
order correction terms would bear no considerable effect on the
results as all the properties are restricted to the case of small
$q'$.

\begin{figure}
\begin{center}
\epsfig{figure=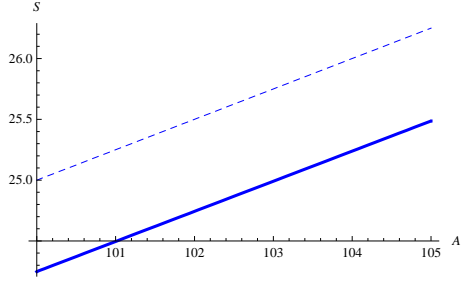,width=6cm}
\end{center}
\caption{\footnotesize Entropy of the brane-world black hole plotted
as a function of the area for $\alpha=-0.2$, $q'=0.1$ and $L_p=1$.
The thick line represents the black hole entropy using MDR and the
dashed line is that of the Bekenstein-Hawking entropy-area
relation.}
\end{figure}

\begin{figure}
\begin{center}
\epsfig{figure=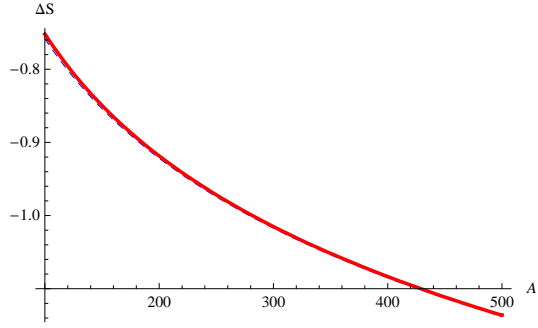,width=7cm}
\end{center}
\caption{\footnotesize Variation of $ \Delta S=S-S_{BH}$ as a
function of the area represented by the thick line, for
$\alpha=-0.2$, $q'=0.1$ and $L_p=1$. The dashed line represents the
logarithmic correction term and is the dominant term. }
\end{figure}

\begin{figure}
\begin{center}
\epsfig{figure=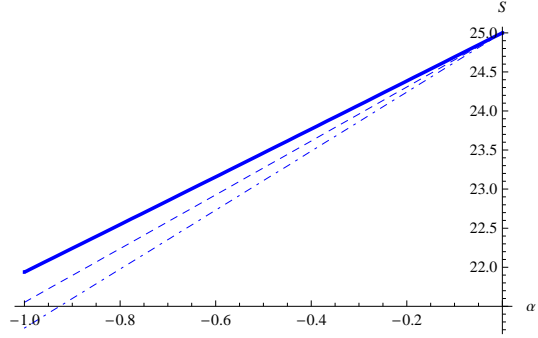,width=7cm}
\end{center}
\caption{\footnotesize Relation between the black hole entropy and
$\alpha$ for $A=100$ and $L_p=1$. The dot-dashed, dashed and thick
lines represent $q'=0$, $q'=1$ and $q'=2$ respectively. }
\end{figure}

\begin{figure}
\begin{center}
\epsfig{figure=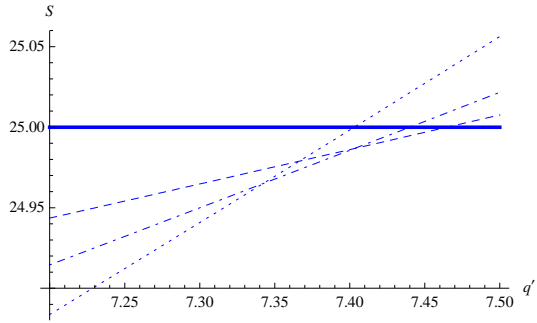,width=7cm}
\end{center}
\caption{\footnotesize Variation of entropy as a function of $q'$
(tidal charge) for different values of $\alpha$ for $A=100$ and
$L_p=1$. The thick, dashed, dot-dashed and dotted lines represent
$\alpha=0$, $\alpha=-0.3$, $\alpha=-0.5$ and $\alpha=-0.8$
respectively. }
\end{figure}

Fig. 1 shows variation of a black hole entropy as a function of its
horizon. One  finds that the predicted entropy within our formalism
is smaller than that of the standard Bekenstein-Hawking. As in our
earlier work \cite{11}, we set the parameter $\alpha$ as a negative
quantity of order one. There, we compared the results of two
approaches, the generalized uncertainty principle  and modified
dispersion relation  within the context of black hole thermodynamics
with that of the string theory and Loop quantum gravity. Demanding
the same results in all approaches and considering string theory and
loop quantum gravity as more comprehensive, we put some constraints
on the form of GUP and MDR. Also, we found that GUP and MDR are not
independent concepts. In fact, they could be equivalent in an
ultimate quantum gravity theory. The existence of a positive minimal
observable length necessitates a positive value for the model
dependent parameter $\alpha$ in the form of GUP. Since we know the
relation between the model dependent parameters in GUP and MDR in
\cite{11}, we set the parameter $\alpha$ as a negative value for MDR
in this paper. It is interesting to note that the existence of a
logarithmic term in the entropy-area relation is necessary within
the present formalism. It would result in a better insight when
dealing and formulating quantum gravity.

The other interesting point is related to the behavior of the
variation of $\Delta S=S-S_{BH}$ which is plotted as a function of
the area in Fig. 2 as a thick line where $S_{BH}$ is the standard
Bekenstein-Hawking entropy. Obviously, the correction terms increase
with area, in spite of the decreasing velocity. Comparison of the
$\Delta S$ curve with that which only contains the logarithmic
correction term (the first correction term) in $S$, represented by a
dashed line in Fig. 2, leads one to the conclusion that the
logarithmic term dominates the correction terms when $q'$ is small.
In other words, the logarithmic term is the dominant correction
term.

The impact of parameter $\alpha$ on the entropy of a black hole with
negative tidal charge is shown in Fig. 3. It is clear that the
entropy decreases with $\alpha$, regardless of the $q'$ magnitude.
However, $q'$ affects the relation between entropy and $\alpha$. The
larger the $q'$ the slower the rate of entropy decrease. In fact,
the tidal charge has information about extra dimensions of space
time and large $q'$ refers to strong gravitational field in the
bulk. As a result one may find that the gravitational field in the
bulk increases the entropy of the black hole on the brane. This
effect can also be seen in Fig.4 where entropy increases as the
tidal charge increases, regardless of the value of $\alpha$. One may
find that in any quantum gravity theory with different $\alpha$, a
gravitational field in the bulk would increase the black hole
entropy on the brane. It may be due to the free propagation of
gravity in the bulk in brane-world scenarios. Since the black hole
entropy is closely related to the gravitational field through the
area, one may conclude that the brane black hole entropy increases
as $q'$ (the tidal charge which is related to bulk gravitational
field) increases. It is clear from the definition of $r_+$ that
negative tidal charge will enlarge the brane black hole horizon.
Thus it is reasonable for the black hole entropy to be influenced by
the effects of the gravitational field in the bulk.

An Estimate of the value of $\alpha$ would now be in order. From the
LQG point of view, the coefficient of the logarithmic term is
$-\frac{1}{2}$. One  then finds the value  $-\frac{4}{3\pi}$ for
$\alpha$. Certainly the exact value of $\alpha$ is still unknown.
Nonetheless, the existence of the logarithmic correction term is
demonstrated in the current quantum-corrected black hole entropy.
Overall, one can compare the results in our work with the results in
\cite{16} which are based on the generalized uncertainty principle
formalism to find that MDR and GUP are equivalent and that they
yield the same results for the brane-world black hole entropy.

It is worth reiterating at this point that one may also have to
impose certain constraints on the general form of GUP according to
\cite{11}, as was discussed earlier, in order to obtain consistent
results. If one tries to investigate the brane-world black hole
entropy from a GUP point of view with higher order correction terms,
the constraints on the general form of the GUP in \cite{11} must be
included. Of course, the results would be functionally consistent
with the results of \cite{16}, but with different coefficients for
similar terms.

\section{Conclusions}
In this paper we have studied the MDR corrections to the brane-world
black hole entropy. We found a good estimate for the value of
$\alpha$ using the LQG approach. The existence of a logarithmic
correction term in our approach would be helpful in providing an
outlook when studying quantum gravity. The quantum-correction
entropy as a function of the horizon area is smaller than the
standard Bekenstein-Hawking one. However, what is the effect of pure
correction terms on the entropy formula? It is clear that the
contribution of the correction terms in the entropy increases with
the horizon area in spite of the decreasing velocity. By taking the
effects of the correction term only, one finds that it behaves as
minus of a logarithmic function. Then the most effective and
important correction term in the entropy formula is the logarithmic
term and irrespective of the value of the tidal charge, the black
hole entropy decreases with $\alpha$. However, it is clear that a
larger charge causes a more slower decrease of  entropy with
$\alpha$. Taking into account that the tidal charge is relevant to
the gravitational field in the bulk, we may conclude that the
gravitational field in the bulk increases the entropy of the black
hole on the brane. As Fig. 4 shows, entropy increases with tidal
charge in quantum gravity models with different $\alpha$.

It is necessary to point that all the above properties are relevant
to the case of small $q'$ which is considered in this paper. It is
easy to show that the results are also valid when the entropy
formula includes higher order correction terms; the contribution of
such terms with increasing powers of $L_p^2/A$  become invariably
negligible.

\end{document}